\documentclass[12pt]{article}
\usepackage{amssymb}
\usepackage{amsmath}

\newcommand{\be}{\begin{eqnarray}}
\newcommand{\ee}{\end{eqnarray}}
\newcommand{\non}{\nonumber}

\newcommand{\tr}{\mathop{\rm tr}\nolimits}
\newcommand{\diag}{\mathop{\rm diag}\nolimits}

\begin{document}

\begin{titlepage}
\strut\hfill UMTG--272
\vspace{.5in}
\begin{center}

\LARGE Revisiting the $Y=0$ open spin chain at one loop\\
\vspace{1in}
\large Rafael I. Nepomechie \footnote{nepomechie@physics.miami.edu}\\[0.8in]
\large Physics Department, P.O. Box 248046, University of Miami\\[0.2in]  
\large Coral Gables, FL 33124 USA\\

\end{center}

\vspace{.5in}

\begin{abstract}
In 2005, Berenstein and V\'azquez determined an open spin chain
Hamiltonian describing the one-loop anomalous dimensions of
determinant-like operators corresponding to open strings attached to
$Y=0$ maximal giant gravitons.  We construct the transfer matrix
(generating functional of conserved quantities) containing this
Hamiltonian, thereby directly proving its integrability.  We find the
eigenvalues of this transfer matrix and the corresponding Bethe
equations, which we compare with proposed all-loop Bethe equations.
We note that the Bethe ansatz solution has a certain ``gauge''
freedom, and is not completely unique.
\end{abstract}

\end{titlepage}

\setcounter{footnote}{0}

\section{Introduction}\label{sec:intro}

The discovery and exploitation of integrability in planar AdS/CFT has
already led to many remarkable results \cite{Beisert:2010jr}, and may
even ultimately lead to the solution of planar ${\cal N}=4$
supersymmetric Yang-Mills (SYM), which is widely regarded as the
``harmonic oscillator'' of 4-dimensional gauge theories.  Much of the
focus has been on the problem of computing the anomalous dimensions of
single-trace operators of ${\cal N}=4$ SYM, which can be mapped to the
problem of determining the eigenvalues of certain integrable {\em
closed} spin-chain Hamiltonians, as observed in the seminal work of
Minahan and Zarembo \cite{Minahan:2002ve}.  \footnote{We have in mind here
``long'' operators.  For operators of finite length, there are
finite-size corrections \cite{Sieg:2010jt, Janik:2010kd,
Bajnok:2010ke}.}
However, progress has also
been made on the problem of computing the anomalous dimensions of
determinant-like operators, which can be mapped to the problem of
determining the eigenvalues of certain {\em open} spin-chain
Hamiltonians \cite{Zoubos:2010kh}.
By the AdS/CFT correspondence, these two types of operators correspond
to states of closed strings and open strings attached to D-branes,
respectively. 

The simplest and most-studied open string/chain example is the
so-called $Y=0$ maximal giant graviton brane \cite{Hofman:2007xp}.
The one-loop $SO(6)$ scalar sector open spin chain Hamiltonian was found by
Berenstein and V\'azquez \cite{Berenstein:2005vf}.  They also
determined the (one-loop) boundary S-matrix, and showed that it
satisfies the boundary Yang-Baxter equation (BYBE)
\cite{Cherednik:1985vs, Sklyanin:1988yz, Ghoshal:1993tm}.  A breakdown
of integrability in the $SU(2)$ subsector at two loops was suspected
in \cite{Agarwal:2006gc, Okamura:2006zr}.  However, Hofman and
Maldacena \cite{Hofman:2007xp} subsequently showed that integrability
is in fact preserved at two loops; and, based on $SU(1|2)$ symmetry,
they proposed an all-loop boundary S-matrix that satisfies the BYBE.
(See also \cite{Ahn:2008df}.)  The scalar factor of the all-loop
boundary S-matrix was proposed in \cite{Chen:2007ec}.  Corresponding
all-loop Bethe equations were proposed by Galleas in
\cite{Galleas:2009ye}.  The bulk worldsheet Yangian symmetry discovered in 
\cite{Beisert:2007ds}, suitably generalized to the case of boundary
scattering \cite{Ahn:2010xa, MacKay:2010ey}, was used to construct
all-loop bound-state boundary S-matrices \cite{Ahn:2010xa,
Palla:2011eu}. Finite-size corrections have been considered in 
\cite{Correa:2009mz, Bajnok:2010ui}. The classical integrability of the 
corresponding string sigma model with boundary has been investigated in 
\cite{Mann:2006rh, Dekel:2011ja}.

Despite the availability of all-loop results, there are still some
important unresolved problems at one loop (where results are generally
most solid): the transfer matrix (generating functional of conserved
quantities) containing the Berenstein-V\'azquez Hamiltonian has not
been constructed, and the corresponding eigenvalues and Bethe
equations have not been determined.  For the corresponding one-loop
$SO(6)$ scalar sector closed spin chain, such results were already
obtained in the original Minahan-Zarembo work \cite{Minahan:2002ve}.

The purpose of our paper is to fill this gap: namely, to construct the
one-loop open-chain transfer matrix, and to determine its eigenvalues.
In this way, we directly prove the integrability of the one-loop
Hamiltonian  \cite{Berenstein:2005vf}, and test the all-loop Bethe equations
\cite{Galleas:2009ye}.  The simpler case of the $SU(3)$ subsector was
treated in \cite{Nepomechie:2009zi, Nepomechie:2009en}. 

The outline of this paper is as follows.  In Section
\ref{sec:transfer}, we recall the open spin chain Hamiltonian found in
\cite{Berenstein:2005vf}, and construct the corresponding transfer
matrix.  In Section \ref{sec:ABA} we use the analytical Bethe ansatz
to determine the eigenvalues of this transfer matrix and the
corresponding Bethe equations.  Surprisingly, the Bethe ansatz
solution has a certain ``gauge'' freedom, and is not completely
unique.  One of the sets of one-loop Bethe equations that we find is
consistent with the all-loop equations.  We conclude in Section
\ref{sec:discussion} with a brief discussion of our results.

\section{Construction of the transfer matrix}\label{sec:transfer}

In \cite{Berenstein:2005vf} Berenstein and V\'azquez identified an open
$SO(6)$ spin chain that describes the one-loop anomalous dimensions of 
determinant-like operators corresponding to open strings attached to
$Y=0$ maximal giant gravitons.  The space of states
is \footnote{Following \cite{Hofman:2007xp}, we define the origin of the spin chain at site 0 
instead of site 1.}
\be
\stackrel{\stackrel{0}{\downarrow}}{C^{5}}  \otimes 
\stackrel{\stackrel{1}{\downarrow}}{C^{6}}  \otimes \cdots
\stackrel{\stackrel{L}{\downarrow}}{C^{6}}  \otimes 
\stackrel{\stackrel{L+1}{\downarrow}}{C^{5}} \,.
\label{Hilbertspace}
\ee
That is, the vector spaces of the ``bulk'' sites (labeled $1, \ldots, 
L$) all have dimension 6, while the vector spaces of the left and right 
``boundary'' sites (labeled 0 and $L+1$, respectively) have dimension 
5. The Hamiltonian is given in Eq.  (2.15), which we rewrite as
\be
H =  Q_{0}^{Y} h_{0, 1} Q_{0}^{Y} + (I - Q_{0}^{\bar Y}) +\sum_{l=1}^{L-1} h_{l, l+1} + 
Q_{L+1}^{Y} h_{L, L+1} Q_{L+1}^{Y} + (I - Q_{L+1}^{\bar Y})  \,,
\label{Hamiltonian}
\ee
where $h_{l,l+1}$ is the bulk two-site Hamiltonian
\be
h_{l,l+1} = \frac{1}{2} {\cal K}_{l,l+1} + I_{l,l+1} -  {\cal 
P}_{l,l+1}  \,.
\label{bulk}
\ee
We have relabeled the sites to run from 0 to $L+1$ (instead of 1 to 
$L$); we have relabeled $Z$ and $\bar Z$ by  $Y$ and $\bar Y$, 
respectively; and for simplicity, we have set the coupling constant $\lambda
\equiv 1$. We note that $Q^{\phi}$ is the projector
\be
Q^{\phi} |\phi \rangle = 0 \,, \quad Q^{\phi} |\varphi \rangle = 
|\varphi \rangle \mbox{  for  } \varphi \ne \phi \,.
\ee 

In the standard basis $|a \rangle = e_{a}$ with $a = 1, \ldots, 6$ 
(elementary $6$-dimensional vectors with
components $\big[e_{a}\big]_{i} = \delta_{a, i}$), the matrices ${\cal P}$ and ${\cal K}$ 
are given by
\be
{\cal P} = \sum_{a,b =1}^{6} e_{a b} \otimes e_{b a} \,, \qquad
{\cal K} = \sum_{a,b =1}^{6} e_{a b} \otimes e_{a b} \,,
\ee
where $e_{a b}$ is the standard elementary $6 \times 6$ matrix whose 
$(a, b)$ matrix element is 1, and all others are zero; i.e., 
$\big[e_{a b}\big]_{ij} = \delta_{a i} \delta_{b j}$.

It is convenient to change to a new basis,
\be
|\tilde  1\rangle &=& |W \rangle = \frac{1}{\sqrt{2}}(e_{1}+i e_{2}) 
\,, \qquad 
|\tilde 2\rangle =|\bar W \rangle = \frac{1}{\sqrt{2}}(e_{1}-i e_{2}) \,, \non \\
|\tilde 3\rangle &=&|Z \rangle = \frac{1}{\sqrt{2}}(e_{3}+i e_{4}) 
\,,  \qquad 
|\tilde 4\rangle =|\bar Z \rangle = \frac{1}{\sqrt{2}}(e_{3}-i e_{4}) \,, \non \\
|\tilde 5\rangle &=&|Y \rangle = \frac{1}{\sqrt{2}}(e_{5}+i e_{6}) 
\,, \qquad 
|\tilde 6\rangle =|\bar Y \rangle = \frac{1}{\sqrt{2}}(e_{5}-i e_{6}) \,.
\label{newbasis}
\ee
Let $U$ be the unitary operator which implements this change of basis,
\be
| \tilde a \rangle = U | a \rangle \,.
\ee
Its matrix elements in the original basis $U_{ba} = 
\langle b| U |a \rangle = \langle b| \tilde a \rangle$
are given by 
\be
U = \frac{1}{\sqrt{2}}
\left( \begin{array}{cccccc}
    1  & 1  \\
    i  & -i  \\
    &  &  1 & 1\\
    &  &  i & -i\\
    &  &   &   &  1 & 1\\
    &  &   &   &  i & -i
\end{array} \right) \,,
\ee
where matrix elements that are zero are left empty.

In this new basis, ${\cal P}$ does not change, but ${\cal K}$ does 
change:
\be
\tilde {\cal P} &=& (U^{\dagger} \otimes U^{\dagger} ) {\cal P} (U 
\otimes U ) = {\cal P} \,, \non \\
\tilde {\cal K} &=& (U^{\dagger} \otimes U^{\dagger} ) {\cal K} (U 
\otimes U ) \,.
\ee
Moreover, in this new basis, the projectors $Q^{Y}$ and $Q^{\bar Y}$ 
are given by the diagonal matrices
\be
Q^{Y} &=& \diag(1,1,1,1,0,1) = 1 - |Y\rangle \langle Y| \,, \non \\
Q^{\bar Y} &=& \diag(1,1,1,1,1,0) = 1 - |\bar Y\rangle \langle \bar Y|\,. 
\label{projectors}
\ee
We therefore arrive at the following explicit matrix representation of the
Hamiltonian 
\be
H = H^{L}_{bt} + \sum_{l=1}^{L-1} \tilde h_{l, l+1} + H^{R}_{bt} \,,
\label{newHamiltonian}
\ee
where the bulk two-site Hamiltonian is given by
\be
\tilde h_{l,l+1} = \frac{1}{2} \tilde {\cal K}_{l,l+1} + I_{l,l+1} -  {\cal 
P}_{l,l+1}  \,,
\label{bulknew}
\ee
and the boundary terms are given by 
\be
H^{L}_{bt} &=& Q_{0}^{Y} \tilde h_{0, 1} Q_{0}^{Y} + (I - Q_{0}^{\bar 
Y}) \,, \label{lbt} \\
H^{R}_{bt} &=& Q_{L+1}^{Y} \tilde h_{L, L+1} Q_{L+1}^{Y} 
+ (I - Q_{L+1}^{\bar Y})\,. \label{rbt}
\ee 
The boundary terms have the property
\be
H^{L}_{bt} = Q_{0}^{Y}\,   H^{L}_{bt}\,  Q_{0}^{Y}\,, \qquad
H^{R}_{bt} = Q_{L+1}^{Y}\, H^{R}_{bt}\, Q_{L+1}^{Y} \,.
\ee
We drop the null rows and columns in these matrices that are due 
to the $Q^{Y}$ projectors. Hence, $H^{L}_{bt}$ and $H^{R}_{bt}$ 
should be understood as $30 \times 30$ matrices acting on $C^{5} 
\times C^{6}$ and $C^{6} \times C^{5}$, respectively, as indicated in 
(\ref{Hilbertspace}).

We now address the problem of constructing the corresponding transfer
matrix that contains this Hamiltonian as well as the higher conserved
charges.  According to Sklyanin \cite{Sklyanin:1988yz}, in order to
construct an open-chain transfer matrix, we need an $R$-matrix that
gives the bulk two-site Hamiltonian; and also left and right
$K$-matrices, which give the left and right boundary terms, 
respectively. \footnote{In order to avoid confusion, it may be worth 
noting that these $R$ and $K$ matrices, even though they satisfy bulk 
and boundary Yang-Baxter equations, have no direct relation to the bulk 
and boundary S-matrices discussed in the Introduction. This fact  
is evident in the more familiar case of the ferromagnetic spin-1/2 
XXX Heisenberg chain: while the bulk S-matrix is a $U(1)$ phase, the 
$R$-matrix is an $SU(2)$-invariant $4 \times 4$ matrix.}

\subsection{$R(u)$ matrix}

We recall that the Yang-Baxter equation is given by
\be
R_{12}(u_{1}-u_{2})\, R_{13}(u_{1})\, R_{23}(u_{2}) =
R_{23}(u_{2})\, R_{13}(u_{1})\, R_{12}(u_{1}-u_{2}) \,.
\label{YBE}
\ee 
An $SO(6)$-invariant solution which acts on $C^{6} \otimes 
C^{6}$ is given (in the new basis) by \cite{Zamolodchikov:1977nu, Minahan:2002ve}
\be
R(u) = \frac{1}{n-2}\left[ u(2u+2-n) I -(2u + 2 - n) {\cal P} + 2u 
\tilde {\cal K} \right] 
\,,
\label{bulkR}
\ee 
with $n=6$. This $R$-matrix indeed produces the bulk two-site
Hamiltonian (\ref{bulknew}), since
\be
\tilde h_{l,l+1} = {\cal P}_{l,l+1} R'_{l,l+1}(0)+\frac{3}{2}I_{l,l+1} \,,
\ee
where the prime denotes differentiation with respect to the spectral 
parameter $u$.

\subsection{$K^{-}(u)$ matrix}

The right $K$-matrix should give the right boundary term (\ref{rbt}) in the 
Hamiltonian.
The $K$-matrix must therefore be operator-valued, rather than 
$c$-number valued. Indeed, the $K$-matrix $K^{-}(u)$ (which acts on $C^{6} \otimes 
C^{5}$)  must satisfy the right ``operator'' BYBE
\be
\lefteqn{R_{12}(u_{1}- u_{2})\, K^{-}_{13}(u_{1})\, R_{12}(u_{1}+u_{2})\, 
K^{-}_{23}(u_{2}) }\non \\
& & = K^{-}_{23}(u_{2})\, R_{12}(u_{1}+u_{2})\, K^{-}_{13}(u_{1})\,R_{12}(u_{1}- 
u_{2}) \,,
\label{RBYBE}
\ee 
where the $R$-matrix is given by (\ref{bulkR}).

We claim that the needed $K$-matrix is given by
\be
K^{-}_{12}(u) = Q^{Y}_{2} R_{12}(u) \tilde K_{1}(u)  R_{12}(u)  
Q^{Y}_{2} \,,
\label{kM}
\ee
where $\tilde K(u)$ is the $6 \times 6$ diagonal $K$-matrix 
\cite{LimaSantos:2003hx, Abdalla:2008kf}
\be
\tilde K(u) = \diag(1-u,1-u,1-u,1-u,u-1,u+1) \,,
\ee
which satisfies the standard (not operator) BYBE
\be
\lefteqn{R_{12}(u_{1}- u_{2})\, \tilde K_{1}(u_{1})\, R_{12}(u_{1}+u_{2})\, 
\tilde K_{2}(u_{2}) }\non \\
& & = \tilde K_{2}(u_{2})\, R_{12}(u_{1}+u_{2})\, \tilde K_{1}(u_{1})\,R_{12}(u_{1}- 
u_{2}) \,.
\ee 
We note that operator $K$-matrices of the general type (\ref{kM}) were introduced
by Frahm and Slavnov \cite{Frahm:1998}, who called them ``projected''
$K$-matrices. The fact that $K$-matrices of this type are needed to 
construct the $SU(3)$ subsector of the $Y=0$ spin chain was noted in \cite{Nepomechie:2009en}.
 
Indeed, we have verified that (\ref{kM}) does satisfy the operator 
BYBE (\ref{RBYBE}), and produces the right boundary term (\ref{rbt}), 
\be
H^{R}_{bt}  = \frac{1}{2}K^{-\ '}(0)  + 2 I \,.
\ee
We also note that 
\be
K^{-}(0) = I \,.
\ee

\subsection{$K^{+}(u)$ matrix}

The left $K$-matrix should give the left boundary term (\ref{lbt}) in the 
Hamiltonian.
For the $SU(3)$ case \cite{Nepomechie:2009zi} and in fact for the 
general $SU(N)$ case, 
we found in \cite{Nepomechie:2009en} that the needed left $K$-matrix can be obtained 
from the right $K$-matrix using Sklyanin's \cite{Sklyanin:1988yz} ``less obvious'' 
isomorphism \footnote{While in \cite{Nepomechie:2009zi} the left 
$K$-matrix is defined to act on $V_{0} V_{1}$, here we define the left 
$K$-matrix to act on $V_{1} V_{0}$; i.e., the two $K$-matrices are 
related simply by permutation of the vector spaces $V_{0}$ and $V_{1}$. The 
isomorphism (\ref{isomorphism}) evidently gives directly the latter form.}
\be
K^{+}_{13}(u) = k(u) \tr_{2} {\cal P}_{12} R_{12}(-2u-\eta) K^{-}_{23}(u) \,,
\label{isomorphism}
\ee
where $\eta$ appears in the crossing-unitarity relation
\be
R_{12}(u)^{t_{1}} \, R_{12}(-u-\eta)^{t_{1}} \propto I \,,
\label{crossingunitarity}
\ee 
and $k(u)$ is an arbitrary scalar factor.

Fortunately, the same trick works also for the $SO(6)$ case. Indeed, the 
$SO(6)$  $R$-matrix (\ref{bulkR}) satisfies the crossing-unitarity property (\ref{crossingunitarity})
with $\eta=-4$. We have verified that the 
$K$-matrix $K^{+}(u)$ given by the isomorphism (\ref{isomorphism}), which acts on $C^{6} \otimes 
C^{5}$, satisfies the left operator BYBE
\be
\lefteqn{R_{12}(-u_{1}+ u_{2})\, K^{+}_{13}(u_{1})^{t_{1}}\, 
R_{12}(-u_{1}-u_{2}-\eta)\, 
K^{+}_{23}(u_{2})^{t_{2}} }\non \\
& & = K^{+}_{23}(u_{2})^{t_{2}}\, R_{12}(-u_{1}-u_{2}-\eta)\, 
K^{+}_{13}(u_{1})^{t_{1}}\,R_{12}(-u_{1} +u_{2}) \,.
\label{LBYBE}
\ee 
And, most importantly, this $K$-matrix produces the left boundary term
(\ref{lbt}),
\be
H^{L}_{bt}  = -\frac{1}{6}\left[ \tr_{a} K^{+}_{a 0}(0) \left( 
R'_{a1}(0)  {\cal P}_{a1} + {\cal P}_{a1} R'_{a1}(0) \right) 
+ \tr_{a} K^{+\ '}_{a 0}(0)  \right] + 
\frac{7}{3} I 
\,,
\ee
where the trace is over a 6-dimensional auxiliary space, which is
discussed further below.  We have fixed the scalar factor in
(\ref{isomorphism}) to be $k(u)=[u(u-2)(2u-1)]^{-1}$ in order to
cancel corresponding terms that appear in the numerator. Finally, we 
note that 
\be
\tr_{a} K^{+}_{a 0}(0) = -3 I \,.
\ee

\subsection{Transfer matrix}

Having gathered all the necessary ingredients, we are now ready to 
assemble them to form the transfer matrix. We introduce a 6-dimensional 
auxiliary, denoted by $a$, and consider operators on the enlarged 
vector space (cf. (\ref{Hilbertspace}))
\be
\stackrel{\stackrel{a}{\downarrow}}{C^{6}}  \otimes 
\stackrel{\stackrel{0}{\downarrow}}{C^{5}}  \otimes 
\stackrel{\stackrel{1}{\downarrow}}{C^{6}}  \otimes \cdots
\stackrel{\stackrel{L}{\downarrow}}{C^{6}}  \otimes 
\stackrel{\stackrel{L+1}{\downarrow}}{C^{5}} \,.
\label{enlargedtspace}
\ee
We define the monodromy matrices $T$ and $\hat T$ by
\be
T_{a 1 \cdots L}(u) = R_{a 1}(u) \cdots R_{a L}(u) \,, \qquad
\hat T_{a 1 \cdots L}(u) = R_{a L}(u) \cdots R_{a 1}(u) \,,
\label{monodromy}
\ee 
where $R(u)$ is given by (\ref{bulkR}).
The transfer matrix is given by \cite{Sklyanin:1988yz}
\be
t(u) = \tr_{a} K^{+}_{a 0}(u)\, T_{a 1 \cdots L}(u) \, K^{-}_{a L+1}(u)\,
\hat T_{a 1 \cdots L}(u)   \,,
\label{transfer}
\ee
where $K^{-}(u)$ and $K^{+}(u)$ are given by (\ref{kM}) and  (\ref{isomorphism}), 
respectively.

By construction, the transfer matrix has the fundamental commutativity property
\be
\left[ t(u) \,, t(v) \right] = 0 \,,
\label{commutativity}
\ee
and contains the Hamiltonian (\ref{newHamiltonian}),
\be
H = -\frac{1}{6} t'(0) + \left(\frac{3}{2}L+\frac{17}{6}\right) I  \,.
\label{Ht}
\ee
The relations (\ref{commutativity}) and (\ref{Ht}) directly imply the
integrability of the Hamiltonian.  Higher conserved quantities can be
obtained from higher derivatives of the transfer matrix at $u=0$.

We note that the transfer matrix $t(u)$ is crossing invariant up to a 
scalar factor,
\be
t(2-u) = \frac{u}{2-u} t(u) \,.
\label{crossing}
\ee
Equivalently, defining a rescaled transfer matrix $\bar t(u)$ by
\be
\bar t(u) = u\, t(u) \,,
\ee
we see that this rescaled transfer matrix is exactly crossing 
invariant,
\be
\bar t(2-u) = \bar t(u) \,.
\ee 
The Hamiltonian is evidently related to the {\em second} derivative of $\bar t(u)$ at 
$u=0$.

\section{Analytical Bethe ansatz}\label{sec:ABA}

The commutativity property (\ref{commutativity}) implies that it is 
possible to find eigenstates $| \Lambda \rangle$ of the transfer 
matrix $t(u)$ (\ref{transfer}) which are independent of $u$,
\be
t(u)\, | \Lambda \rangle = \Lambda(u)\, | \Lambda \rangle \,.
\ee
We turn now to the problem of determining the eigenvalues
$\Lambda(u)$.  We proceed by the analytical Bethe ansatz approach,
along the lines in \cite{Minahan:2002ve, Nepomechie:2009zi}.  We
choose as a reference state
\be
|0\rangle = {\small \left( \begin{array}{c}
1 \\
0 \\
0 \\
0 \\
0 \end{array} \right) \otimes 
\left( \begin{array}{c}
1 \\
0 \\
0 \\
0 \\
0 \\
0 \end{array} \right)^{\otimes L}  \otimes 
\left( \begin{array}{c}
1 \\
0 \\
0 \\
0 \\
0 \end{array} \right)} \,,
\label{vacuum}
\ee 
which is a ground state of the Hamiltonian (\ref{newHamiltonian}) with 
eigenvalue 0. We denote the 
corresponding eigenvalue of the transfer matrix by $\Lambda_{0}(u)$,
\be
t(u)\, |0 \rangle = \Lambda_{0}(u)\, | 0 \rangle \,.
\ee
On the basis of results for $L=0,1,2$, we obtain the following
conjecture for the vacuum eigenvalue
\be
\hspace{-0.2in}\Lambda_{0}(u) &=& \frac{1}{4^{L} d(u)}
\Big[ a(u) (u-2)^{2L} (u-1)^{2L}
+ b(u)  (u-1)^{2L} 
u^{2L}
+ 4 c(u)  (u-2)^{2L} u^{2L} \Big] \,,
\label{vacuumeigenvalue}
\ee
where
\be
a(u) &=& (u-2)^4 (u-1)^5 (2 u-3)^2 \,, \non \\
b(u) &=&  (u-2)  (u-1)^5 u^3 (2 u-1)^2 \,, \non \\
c(u) &=& 4 (u-2)^4 (u-1)^4 u^3\,, \non \\
d(u) &=& 16 (2 u-3) (2 u-1) \,.
\ee 
A general eigenvalue should be a ``dressed'' vacuum 
eigenvalue, \footnote{For the corresponding closed-chain
result, see (4.28) in \cite{Minahan:2002ve} and references therein.}
\be
\Lambda(u) &=& \frac{1}{4^{L} d(u)}
\Big\{ a(u) (u-2)^{2L} (u-1)^{2L} \frac{Q_{1}(u+\frac{1}{2})}{Q_{1}(u-\frac{1}{2})}
+ b(u)  (u-1)^{2L} u^{2L} 
\frac{Q_{1}(u-\frac{5}{2})}{Q_{1}(u-\frac{3}{2})} \non\\
&+& (u-2)^{2L} u^{2L} \Big[ c_{1}(u)\frac{Q_{1}(u-\frac{3}{2}) 
Q_{2}(u) Q_{3}(u)}
{Q_{1}(u-\frac{1}{2})Q_{2}(u-1) Q_{3}(u-1)} \non \\
&+& 
c_{2}(u)\frac{Q_{1}(u-\frac{1}{2}) 
Q_{2}(u-2) Q_{3}(u-2)}
{Q_{1}(u-\frac{3}{2})Q_{2}(u-1) Q_{3}(u-1)} \non \\
&+&  c_{3}(u)\frac{Q_{2}(u) Q_{3}(u-2)}
{Q_{2}(u-1) Q_{3}(u-1)} + 
c_{4}(u)\frac{Q_{2}(u-2) Q_{3}(u)}
{Q_{2}(u-1) Q_{3}(u-1)} \Big]
\Big\}\,,
\label{Lambda}
\ee
where
\be
Q_{a}(u) \equiv \prod_{k=1}^{m_{a}} (u- i u_{a,k}) (u+ i u_{a,k}) \,, \quad a = 1, 2, 3,
\label{Qs}
\ee
which have the property $Q_{a}(-u)=Q_{a}(u)$. The functions 
$c_{1}(u), \ldots, c_{4}(u)$ must satisfy
\be
c_{1}(u) + c_{2}(u) + c_{3}(u) + c_{4}(u) = 4c(u) 
\label{constraint0}\,,
\ee
but are otherwise still to be determined.

The eigenvalues of the Hamiltonian (\ref{newHamiltonian}) now follow from 
(\ref{Ht}) and (\ref{Lambda}),
\be
E = -\frac{1}{6} \Lambda'(0) + \frac{3}{2}L+\frac{17}{6} = 
\sum_{k=1}^{m_{1}}\frac{1}{u_{1,k}^{2}+\frac{1}{4}} \,,
\label{energy}
\ee 
which is the same expression that one obtains for the corresponding closed chain 
\cite{Minahan:2002ve}.

The crossing symmetry (\ref{crossing}) implies a corresponding 
property of the eigenvalues 
\be
\Lambda(2-u) = \frac{u}{2-u} \Lambda(u) \,,
\ee
which in turn implies the constraints
\be
c_{1}(2-u) = \frac{u}{2-u} c_{2}(u) \,, \qquad
c_{3}(2-u) = \frac{u}{2-u} c_{4}(u) 
\label{constraint1} \,.
\ee

The Bethe equations for the zeros $u_{1,k}$ of the function
$Q_{1}(u)$ (\ref{Qs}) follow from the fact that $\Lambda(u)$ in 
(\ref{Lambda}) is analytic at $u=i u_{1,k}+\frac{1}{2}$, which implies
\be
\left(\frac{i u_{1,k}+\frac{1}{2}}{i u_{1,k}-\frac{1}{2}}\right)^{2L} 
= -\frac{a(i u_{1,k}+\frac{1}{2})}{c_{1}(i u_{1,k}+\frac{1}{2})}
\frac{Q_{1}(i u_{1,k}+1)\, Q_{2}(i u_{1,k}-\frac{1}{2})\, Q_{3}(i u_{1,k}-\frac{1}{2})}
{Q_{1}(i u_{1,k}-1)\, Q_{2}(i u_{1,k}+\frac{1}{2})\, Q_{3}(i 
u_{1,k}+\frac{1}{2})} \,.
\label{BAE1}
\ee
We should obtain the same Bethe equations by considering instead the 
poles at $u=i u_{1,k}+\frac{3}{2}$, which implies the constraint
\be
\frac{a(u)}{c_{1}(u)} = \frac{c_{2}(u+1)}{b(u+1)}\,.
\label{constraint2}
\ee
Similarly, by considering the poles at $u=i u_{2,k}+1$, we obtain the 
Bethe equations
\be
1
= -\frac{c_{1}(i u_{2,k}+1)}{c_{4}(i u_{2,k}+1)}
\frac{Q_{2}(i u_{2,k}+1)\, Q_{1}(i u_{2,k}-\frac{1}{2})}
{Q_{2}(i u_{2,k}-1)\, Q_{1}(i u_{2,k}+\frac{1}{2})} \,,
\label{BAE2}
\ee
and the constraint
\be
\frac{c_{1}(u)}{c_{4}(u)} = \frac{c_{3}(u)}{c_{2}(u)}\,.
\label{constraint3}
\ee
Finally, by considering the poles at $u=i u_{3,k}+1$, we obtain the 
Bethe equations
\be
1
= -\frac{c_{1}(i u_{3,k}+1)}{c_{3}(i u_{3,k}+1)}
\frac{Q_{3}(i u_{3,k}+1)\, Q_{1}(i u_{3,k}-\frac{1}{2})}
{Q_{3}(i u_{3,k}-1)\, Q_{1}(i u_{3,k}+\frac{1}{2})} \,,
\label{BAE3}
\ee
and the same constraint (\ref{constraint3}).  It can be shown that the
requirement that the eigenvalues (\ref{Lambda}) be analytic at $u= \frac{1}{2},
\frac{3}{2}$ (where $d(u)$ has zeros) does not lead to further
constraints.  \footnote{It may be possible to derive further
constraints from the requirement that the eigenvalues of the transfer
matrix obtained by fusion in the auxiliary space also be analytic.
However, such calculations would be difficult, and we shall not pursue
them here.}

The constraints (\ref{constraint0}), (\ref{constraint1}),
(\ref{constraint2}), (\ref{constraint3}) do not uniquely determine the
functions $c_{1}(u), \ldots, c_{4}(u)$.  Moreover, the problem of
determining the transfer-matrix eigenvalues $\Lambda(u)$ in the $T-Q$
form (\ref{Lambda}) does not have a unique solution.  Indeed,
consider the following ansatz for these functions
\be
c_{1}(u) &=&  u^{n} (u-1)^{5-n} (u-2)^{4} (2u-3)^{2}\,, \non \\
c_{2}(u) &=&  u^{3} (u-1)^{5-n} (u-2)^{1+n} (2u-1)^{2}\,, \non \\
c_{3}(u) &=&  u^{3+m} (u-1)^{5-n}(u-2)^{1+n-m} (2u-1)(2u-3)\,, \non \\
c_{4}(u) &=&  u^{n-m} (u-1)^{5-n} (u-2)^{4+m} (2u-1)(2u-3)\,,
\label{cs}
\ee 
where $n$ and $m$ are integers.  All the constraints
(\ref{constraint0}), (\ref{constraint1}), (\ref{constraint2}),
(\ref{constraint3}) are then satisfied for the following four sets of
$(n,m)$ values
\be
{\rm case\ I\ \ \, } &:& \quad n=3\,, \quad m = 0 \,, \non \\
{\rm case\ IIa} &:& \quad n=5\,, \quad m = 0 \,, \non \\
{\rm case\ IIb} &:& \quad n=5\,, \quad m = 2 \,, \non \\
{\rm case\ III} &:& \quad n=7\,, \quad m = 2 \,,
\label{cases}
\ee 
which we have designated as cases I, IIa, IIb, III, respectively.

The Bethe equations 
(\ref{BAE1}), (\ref{BAE2}), (\ref{BAE3}) can now be rewritten in the 
more familiar form
\be
e_{1}(u_{1, k})^{2L+n-1} &=& 
\prod_{j=1 \atop j\ne k}^{m_{1}} e_{2}(u_{1, k} - u_{1, j})\, 
e_{2}(u_{1, k} + u_{1, j}) 
\prod_{j=1}^{m_{2}} e_{-1}(u_{1, k} - 
u_{2, j})\, e_{-1}(u_{1, k} + u_{2, j}) \non \\
& & \times \prod_{j=1}^{m_{3}} e_{-1}(u_{1, k} - 
u_{3, j})\, e_{-1}(u_{1, k} + u_{3, j}) \,, \quad k = 1\,, \ldots 
\,, m_{1} \,, \label{BAESO6a} 
\ee
\be 
1 &=&  e_{2}(u_{2, k})^{m}
\prod_{j=1 \atop j\ne k}^{m_{2}} e_{2}(u_{2, k} - u_{2, j})\, 
e_{2}(u_{2, k} + u_{2, j})\non \\
& & \times  \prod_{j=1}^{m_{1}} e_{-1}(u_{2, k} - 
u_{1, j})\, e_{-1}(u_{2, k} + u_{1, j}) \,,  \quad k = 1\,, \ldots 
\,, m_{2} \,, \label{BAESO6b}
\ee
\be 
1 &=&  e_{2}(u_{3, k})^{n-m-3}
\prod_{j=1 \atop j\ne k}^{m_{3}} e_{2}(u_{3, k} - u_{3, j})\, 
e_{2}(u_{3, k} + u_{3, j})\non \\
& & \times  \prod_{j=1}^{m_{1}} e_{-1}(u_{3, k} - 
u_{1, j})\, e_{-1}(u_{3, k} + u_{1, j}) \,,  \quad k = 1\,, \ldots 
\,, m_{3} \,, \label{BAESO6c}
\ee 
where
\be
e_{n}(u) = \frac{u +  \frac{i n}{2}}{u -  \frac{i n}{2}} \,.
\label{notation}
\ee
For cases I and III, the Bethe equations are  
symmetric under the interchange of Bethe roots of types 2 and 3 
(i.e., $u_{2, i} \leftrightarrow u_{3, i}$), 
as in the closed-chain case \cite{Minahan:2002ve}. The Bethe equations
for cases IIa and IIb transform into each other under the interchange
of Bethe roots of types 2 and 3.

At least for small values of $L$, the Bethe ansatz solutions
corresponding to each of the cases (\ref{cases}) are complete; i.e., they
reproduce all the transfer-matrix eigenvalues, of which there are $6^{L} 5^{2}$
according to (\ref{Hilbertspace}).  Indeed, following the approach
described in Appendix B of \cite{Nepomechie:2009zi}, we have verified
the completeness for $L=0$ and $L=1$, for which the total number of
states is 25 and 150, respectively. The results for case I are
summarized in Table \ref{table:L0} and Table \ref{table:L1},
respectively.  For cases II and III, the Bethe roots of type 1 (i.e.,
$\{ u_{1, k} \}$) describing a given eigenvalue are the same as for
case I, but the Bethe roots of types 2 and 3 are 
different.\footnote{Generally, there are fewer Bethe roots of types 2 and 3
for cases II and III in comparison with case I.}  This does
not lead to differences in the energy (\ref{energy}) or higher conserved
quantities, which depend only on the Bethe roots of type 1. Hence,
the various cases can be regarded as equivalent. Gauge-like 
transformations relating these cases are discussed in Section 
\ref{sec:cases}.

For case I, the above one-loop Bethe equations 
coincide with those that we previously conjectured, and obtained
from the all-loop Bethe equations \cite{Galleas:2009ye} by performing
the weak-coupling limit and then reducing to the $SO(6)$ sector, which
are given in Eqs.  (3.1) and (4.20) in \cite{Nepomechie:2009zi},
respectively. For cases II and III, the Bethe equations are evidently slightly 
different: in (\ref{BAESO6a}) the term $e_{1}(u_{1, k})$ has a different
power; and (\ref{BAESO6b}) and/or (\ref{BAESO6c}) contain an additional
factor $e_{2}^{2}$.

\section{Discussion}\label{sec:discussion}

We have shown that the Berenstein-V\'azquez Hamiltonian
(\ref{newHamiltonian}) is contained in the commuting transfer matrix
(\ref{transfer}), which directly implies the integrability of this
Hamiltonian.  One key ingredient is the left $K$-matrix (\ref{kM}), which is
of the projected type \cite{Frahm:1998}. Another key ingredient is 
the right $K$-matrix (\ref{isomorphism}), obtained through a 
seldom-used isomorphism noted in \cite{Sklyanin:1988yz}.
We have also found
expressions (\ref{Lambda}), (\ref{Qs}), (\ref{cs}), (\ref{cases}) for
the eigenvalues of this transfer matrix in terms of roots of the Bethe
equations (\ref{BAESO6a})-(\ref{BAESO6c}).  We have therefore
completed the generalization of important closed-chain results of
Minahan and Zarembo \cite{Minahan:2002ve} to the $Y=0$ open chain.

We have seen that the Bethe equations based on the vacuum eigenvalue
(\ref{vacuumeigenvalue}) are not uniquely fixed -- a certain ``gauge''
freedom exists.  A similar phenomenon may occur for other integrable
spin chains with symmetry group of rank greater than one.  (This
phenomenon is distinct from the well-known duality transformations of
supersymmetric spin chains, see e.g. \cite{Beisert:2005di}.)  For case
I in (\ref{cases}), the one-loop Bethe equations
(\ref{BAESO6a})-(\ref{BAESO6c}) agree with those obtained in
\cite{Nepomechie:2009zi} from the all-loop Bethe equations
\cite{Galleas:2009ye} by performing the weak-coupling limit and then
reducing to the $SO(6)$ sector. Hence, our results provide support 
for those all-loop Bethe equations.

The analytical Bethe ansatz approach that we followed here is
heuristic, and requires making some assumptions, in particular
(\ref{cs}).  It would be useful to carry out a more rigorous analysis
based on nested algebraic Bethe ansatz.  For the $O(2N)$ closed spin
chain, a nested algebraic Bethe ansatz was developed in
\cite{Martins:1997wb}.  One would need to generalize that approach to
the open chain with projected $K$-matrices considered here.  Besides
the difficulty of working out the necessary commutation relations, one
can foresee the following further difficulty: after the first level of
nesting, the ``bulk'' quantum space will have 4 dimensions (two less
than the original 6 dimensions), while the ``boundary'' quantum spaces
will still have 5 dimensions.  That is, the bulk and boundary quantum
spaces will not have the same dimension.  The reduced transfer matrix
will therefore not be given by the tensor product of two usual
transfer matrices, as happens at the final stage in the $GL(N)$ case
\cite{Nepomechie:2009en}.

We have restricted our attention here to the $Y=0$ maximal giant
graviton brane, which is the simplest known integrable open
string/chain example.  It may be interesting to consider the
corresponding problem in more complicated examples, such as the $Z=0$
maximal giant graviton brane \cite{Hofman:2007xp}.

\section*{Acknowledgments}
I am grateful to Antonio Lima-Santos for valuable correspondence at an 
early stage of this project.
This work was supported in part by the National Science Foundation 
under Grant PHY-0854366.

\appendix

\section{``Gauge'' transformations among the cases}\label{sec:cases}

Let us denote by $c_{i}^{\rm I}(u), c_{i}^{\rm IIa}(u)$, etc.  the
functions (\ref{cs}) for cases I, IIa, etc.  in (\ref{cases}),
respectively.  It is easy to see that the functions for case IIa are
related to the corresponding ones for case I as follows
\be
c_{1}^{\rm IIa}(u) &=& \left(\frac{u}{u-1}\right)^{2}\, c_{1}^{\rm 
I}(u) \,, \non \\
c_{2}^{\rm IIa}(u) &=& \left(\frac{u-2}{u-1}\right)^{2}\, c_{2}^{\rm 
I}(u) \,, \non \\
c_{3}^{\rm IIa}(u) &=& \left(\frac{u-2}{u-1}\right)^{2}\, c_{3}^{\rm 
I}(u) \,, \non \\
c_{4}^{\rm IIa}(u) &=& \left(\frac{u}{u-1}\right)^{2}\, c_{4}^{\rm 
I}(u) \,. \label{IIareltn}
\ee 
Similarly, 
the functions for case IIb are
related to the corresponding ones for case I as follows
\be
c_{1}^{\rm IIb}(u) &=& \left(\frac{u}{u-1}\right)^{2}\, c_{1}^{\rm 
I}(u) \,, \non \\
c_{2}^{\rm IIb}(u) &=& \left(\frac{u-2}{u-1}\right)^{2}\, c_{2}^{\rm 
I}(u) \,, \non \\
c_{3}^{\rm IIb}(u) &=& \left(\frac{u}{u-1}\right)^{2}\, c_{3}^{\rm 
I}(u) \,, \non \\
c_{4}^{\rm IIb}(u) &=& \left(\frac{u-2}{u-1}\right)^{2}\, c_{4}^{\rm 
I}(u) \,. \label{IIbreltn}
\ee 
Finally, 
the functions for case III are
related to the corresponding ones for case I as follows
\be
c_{1}^{\rm III}(u) &=& \left(\frac{u}{u-1}\right)^{4}\, c_{1}^{\rm 
I}(u) \,, \non \\
c_{2}^{\rm III}(u) &=& \left(\frac{u-2}{u-1}\right)^{4}\, c_{2}^{\rm 
I}(u) \,, \non \\
c_{3}^{\rm III}(u) &=& \frac{u^{2} (u-2)^{2}}{(u-1)^{4}}\, c_{3}^{\rm 
I}(u) \,, \non \\
c_{4}^{\rm III}(u) &=& \frac{u^{2} (u-2)^{2}}{(u-1)^{4}}\, c_{4}^{\rm 
I}(u) \,. \label{IIIreltn}
\ee 

Let us now consider the expression (\ref{Lambda}) for $\Lambda(u)$,
and identify the functions $c_{i}(u)$ there with $c_{i}^{\rm I}(u)$.
Our observation is that, by making in that
expression the simple transformation
\be
Q_{3}(u) \rightarrow u^{2} Q_{3}(u) \,,
\label{Q3transf}
\ee
we obtain, in view of (\ref{IIareltn}), the corresponding expression  
for $\Lambda(u)$ in terms of $c_{i}^{\rm IIa}(u)$. Similarly, by making instead the 
transformation
\be
Q_{2}(u) \rightarrow u^{2} Q_{2}(u) \,,
\label{Q2transf}
\ee
we obtain, in view of (\ref{IIbreltn}), the expression for
$\Lambda(u)$ in terms of $c_{i}^{\rm IIb}(u)$.  Finally, 
in view of  (\ref{IIIreltn}), by making both
transformations (\ref{Q3transf}) and (\ref{Q2transf}), we obtain the
expression for $\Lambda(u)$ in terms of $c_{i}^{\rm III}(u)$.

Similarly, let us consider the Bethe equations in their original form
(\ref{BAE1}), (\ref{BAE2}), (\ref{BAE3}), and identify the functions
$c_{i}(u)$ there with $c_{i}^{\rm I}(u)$.  By making the
transformations (\ref{Q3transf}) and/or (\ref{Q2transf}), we obtain the
Bethe equations for the other cases.

In short, (\ref{Q3transf}) and (\ref{Q2transf}) are sorts of
(discrete) ``gauge'' transformations that relate the four cases
(\ref{cases}) of Bethe ansatz solutions.  In our numerical
investigations, we have observed that for case I there are generally
more zero Bethe roots of types 2 and 3 in comparison with cases II and
III, which is consistent with the presence of additional factors of
$u^{2}$ in the corresponding $Q$ functions.

Related transformations were 
briefly discussed in \cite{Nepomechie:2009en}. There, the 
transformations involve also $Q_{1}(u)$, and therefore, relate Bethe 
ansatz solutions based on reference states with different energies.

\begin{table}[h!] 
  \centering
  \begin{tabular}{|c|c|c|c|}\hline
     deg & $\{ u_{1, k} \}$ & $\{ u_{2, k} \}$ & $\{ u_{3, k} \}$ \\
    \hline
     9  &  -- &  --  &  --\\
     6	&  1/2  &  0  &  -- \\
     4  &  $\sqrt{3}/6$  &  0  &  0\\
     4	&  $\sqrt{3}/2$  &  0  &  0 \\
     2	&  $\pm i/2$  &  0  & 0 \\   
    \hline
   \end{tabular}
   \caption{Degeneracy and Bethe 
   roots for case I with $L = 0$.}
   \label{table:L0}
\end{table}
\begin{table}[h!] 
  \centering
  \begin{tabular}{|c|c|c|c|}\hline
     deg & $\{ u_{1, k} \}$ & $\{ u_{2, k} \}$ & $\{ u_{3, k} \}$ \\
    \hline
     16  &  --  &  --  &  -- \\
     16  &   $\sqrt{3}/6$  &  0  & -- \\
     16  &   $\sqrt{3}/2$  &  0  & -- \\
      9  &  1/2  &  --  & -- \\
      9  &  1/2  & 0  & 0 \\
      9  &  $(\sqrt{2} +1)/2$   &  0  & 0 \\
      9  &  $(\sqrt{2} -1)/2$   &  0  & 0 \\ 
      6  &   $\pm i/2$  &  0  & -- \\
     10  &   $\pm i/2$  &  0   & 0 \\
     14  &  0.230955, 0.668326   &  0  & 0 \\
     14  &  $0.716015 \pm 0.512521 i$   & 0    & 0 \\
     14  &   $\sqrt{3}/6, \sqrt{3}/2$  &  $\sqrt{6}/3$  & 0 \\
      4  &   0.415511, 1.15211  &  1  & 1 \\
      1  &   $1/2, \pm i/2$  &  $0, i$  & $0, i$ \\
      1  & 1/2, $0.479716 \pm 0.971633 i$ & $0, 1.38848 i$ & $0, 1.38848 i$  \\
      1  &  1/2, 0.208963, 1.02227 & 0, 0.767271  & 0, 0.767271  \\
      1  &  1/2, $0.414496 \pm 0.502211 i$ & $0, 0.812907 i$   & $0, 0.812907 i$ \\
    \hline
   \end{tabular}
   \caption{Degeneracy and Bethe 
   roots for case I with $L = 1$.}
   \label{table:L1}
\end{table}

\newpage

\providecommand{\href}[2]{#2}\begingroup\raggedright\endgroup

\end{document}